# A novel pattern recognition system for detecting Android malware by analyzing suspicious boot sequences

Jorge Maestre Vidal[a,*], Marco Antonio Sotelo Monge[a], Luis Javier García Villalba[a,*]

[a]*Group of Analysis, Security and Systems (GASS), Department of Software Engineering and Artificial Intelligence (DISIA), School of Computer Science, Office 431, Universidad Complutense de Madrid (UCM), Calle Profesor José García Santesmases s/n, Ciudad Universitaria, 28040 Madrid, Spain*

**Abstract**

In this paper, a malware detection system for smartphones based on the study of the dynamic behavior of suspicious applications is proposed. The main goal is to prevent the installation of the malicious software on the victim systems. Due to its popularity, the approach focuses on the malware addressed against the Android platform. For that purpose, only the system calls performed during the startup of applications are studied. Thereby the amount of information to be considered is reduced, since only activities related with their initialization are taken into account. The proposal defines a pattern recognition system with three processing layers: monitoring, analysis and decision-making. First, in order to extract the sequences of system calls, the potentially compromised applications are executed on a safe and isolated environment. Then the analysis step generates the metrics required for decision-making. This level combines sequence alignment algorithms with bagging , which allow scoring the similarity between the extracted sequences considering their regions of greatest resemblance. At the decision-making stage, the Wilcoxon signed-rank test is implemented. There the new software is labeled as legitimate or malicious. The proposal has been tested in different experiments, among them an in-depth study of a particular use case and its effectiveness when analyzing samples from a well-known public domain dataset. Promising experimental results have been shown, hence demonstrating that this is a good complement to the strategies of the bibliography.

## 1. Introduction

Over recent years a significant growth in the popularity of mobile devices was observed. It was empowered by their large capacity of connectivity, accessibility, and versatility. Consequently, users increasingly rely

*Tel. +34 91 394 76 38, Fax: +34 91 394 75 47

 *Email addresses:* jmaestre@ucm.es (Jorge Maestre Vidal), masotelo@ucm.es (Marco Antonio Sotelo Monge), javiergv@fdi.ucm.es (Luis Javier García Villalba)

on these technologies to perform activities of special sensitivity, such as e-commerce, sharing assets or management of confidential information. This places smartphones directly in the line of fire for cyber criminals, as has warned the European Network and Information Security Agency (ENISA) on their annual report [30]. This publication does not simply estimate the increase of these threats; it also alerts of their sophistication, which make them difficult to be detected by the current defense schemes. In addition it is also important to highlight the risk of the migration of the classical attacks to the mobile platforms, where the adaptation of malware is one of the most common practices. According to the European Police Office (Europol), behind this laborious task, often complex networks of organized crime are hidden [13]. From among their most effective spreading strategies, the publication in application stores, social engineering and exploitation of vulnerabilities in communication protocols are the most frequent. On the other hand, there are several studies pointing that the Android operating system is the main target for the attackers. For example, in [6] it is indicated that 99% of the malware for mobile devices is addressed against this platform. This is because at present, Android is the most widespread framework. Given the lack of effectiveness of the security methods practiced by its distribution markets and the overconfidence of many of the users (often fueled by lack of knowledge), criminals are able to propagate Android malware rapidly and indiscriminately. To combat this threat, the research community has developed different proposals. In [42] many of them are deeply analyzed, and the evolution of the malicious applications is studied. From this paper it is possible to observe the main causes that led to the failure of many of the current approaches, where the limitation on computational resources is one of the most problematic. Consequently, the detection systems tend to have high false positive rates, pose difficulties to operate in real time, incorporate vulnerabilities related with privacy or penalize the quality of service on the devices.

In order to contribute to mitigate these threats, this paper introduces a pattern recognition system for recognition of malware on the application stores. Its main goal is to prevent the installation of the malicious software on the victim system. To do this, applications downloaded are analyzed in a safe and isolated environment, prior to their deployment on real devices. Unlike the other publications, this approach focuses on the sequences of actions involved in the boot of the suspicious programs. Thus the amount of information to be monitored is considerably reduced. Another important contribution is to adopt the sequence alignment methods provided by bioinformatics to assess the similitude between different applications. The sequences of the boot system calls are handled as strings of amino acids, in such way that their alignment allows distinguishing regions of greatest similarity, which facilitate the decision-making on the detection engine. To determine if the difference between the scores of two groups of test sequences is representative, the Wilcoxon signed-rank test is applied. If a significant difference between the analyzed applications with the set of sequences associated to their legitimate execution is detected, then the program is tagged as malicious. This set is constantly updated by information voluntarily shared by legitimate users. The effectiveness of the proposal has been evaluated by testing the approach with malicious and legitimate samples from the collections Genome [52] and Drebin [2].

The paper is divided into seven sections: the first of them is the present introduction. The background necessary for a better understanding of the approach is described in section II. The bases of the sequence alignment are detailed in section III. The proposed intrusion detection system is introduced in section IV. Experiments, datasets and the evaluation methodology are described in section V. Results are discussed in section VI. Finally, conclusions and future work are presented in section VII

## 2. Background

### 2.1. Malware against Android

The current state of the malware for Android systems is rooted in the evolution of the malicious software in mobile devices. The first specimens, like Cabir (2004) or CommWarrior (2005), exploited vulnerabilities on these technologies in order to be spread through basic communication protocols, such as Multimedia Messaging Service (MMS) or Bluetooth [1]. Although they were relatively harmless, they could generate money losses related to their propagation messages. This led the attackers to pretend obtaining economic benefit of the intrusions, prompting the appearance of specimens such as RedBrowser (2005) or Yxes (2009), which leverage the premium-rate SMS services [9]. The latter would be the precursor of the botnets of mobile devices. The adaptation of well-known threats for personal computers to smartphones reaches an important landmark in Zitmo (2010), the mobile version of the banking botnet Zeus [17]. By then Android already occupied a large market niche, which gave rise to the discovery of new malware specific for this operative system. This is the case of Gemini (2010), DroidKungFu (2011) or Plankton (2011). They were distributed both from the official Android application store (Google Play) and from unofficial third party stores [5]. In more recent years, the different organizations for cyber defense warned of a significant growth of this kind of malicious software. Furthermore, new ways to profit these threats are emerging, such as those related with adware, riskware, spyware, etc. A good example is observed in the case of the ransomware. It is a class of malware focused on the extortion of the victims. The ransomware is capable of blocking some system functionalities and ask for money (ransom) in exchange for their release. The first discovered ransomware for Android is FakeDetect (2013), a family of malware that operates imitating antivirus software, and demands a ransom for unlock the system assets [23]. Other examples of ransomware are FakeAV (2013), CryptoLocker (2014), Koler (2014) or Locker (2015) [12].

The malware specific for Android, like any other application developed for this platform, is distributed compacted in the Application PacKage file format (APK). The malicious APK files are mainly distributed in official or third party stores, although there are other less frequent ways, such as social engineering (spear fishing, baiting, etc.) or exploitation of vulnerabilities. Then they acquire the capability of propagation via communication protocols (Bluetooth, Wi-Fi, NTC, etc.) and by different channels (email, instant messaging, social networks, etc.). This paper aims on the interception of malware in the application stores. In order to reduce these threats, the official markets require all applications to be digitally signed and certified before they are installed. Therefore, through hash functions the integrity of the packages is checked. They also implement various intrusion detection methods, such as static and dynamic analysis for the recognition of previously known threats or anomalous contents. But even though a priori these seem a good way to prevent the distribution of malware and legitimate applications previously unpacked and poisoned by attackers, in practice they are not so effective. This is due to various reasons, emphasizing among them the emerging of new vulnerabilities and that the certificates do not need to be signed by a certificate authority or could be forged, as is discussed in [35]. On the other hand, in order to evade such intrusion detection systems, attackers are progressively resorting to obfuscation strategies. The problem on these adversarial attacks is widely discussed in publications like [29], where it is shown that by mutations on the section of the Android executable that contains the infection vector, it is possible to deceive several detection engines

*2.2. Related Works*

In recent years different surveys about the state of the art on security in smartphones have been published. Some of them propose a general overview, as is the case of [42, 26]; others deepen on more specific topics, where the Android operating system is one of the most frequent [14]. Finally, there are surveys focused on specific threats, such as [34], which delve into the problem of malware. Throughout the bibliography, the limitation of the computational resources provided by the mobile devices has determined how the monitored information is processed at the defensive process. One way of committing to this task is to perform the analysis on the devices themselves. This poses the advantage of not relying on external deployments of communications with collaborative systems. In addition, the exploitation of vulnerabilities related with privacy is more difficult [10]. But as stated in [5], they also tend to emission of more false positives, and they are particularly sensitive against adversarial attacks [29]. This is because in order to operate properly, the most accurate methods require more memory, processing power or battery consumption than they have, as discussed in [44]. As an alternative, several approaches delegate the more complex analysis tasks to external services, even assuming the risks that this entails.

Another important aspect with significant relevance to the previous proposals is the decision of the features studied by the intrusion detection systems. This problem is not particularly representative for approaches based on recognition of previously known threats [15]. However, as outlined in [18], when the detection is based on the identification of anomalous behaviors, it is crucial to ensure the success of the selected characteristics. Due to this it is understandable that a large amount of taxonomies considers these features as the main distinction traits between the different anomaly-based analysis methods [16]. On this basis, the proposals can be distinguished into four great groups: analysis considering static features, dynamic, mixed or metadata.

- *Static analysis*. The analysis of static characteristics inspects the applications at their execution time looking for malicious contents. For this purpose different information is extracted, such as binary code, privileges requested, hardware resources or connectivity. A good example of this class is observed in [28], where these features are studied in the exploration of the various application stores, searching particular specimens. In [50] the source code of applications in the Package Manager Service (PMS) of Android for malware able to gain privileges is analyzed. In [19] the intrusion detection is proposed taking into account the privileges requested by the applications. Alternatively, in [2] this distinction is made by querying the *AndroidManifest.xml*, where the hardware to be requested is indicated. In [43] the code structure is analyzed for identifying similarities between the different malware samples. With this, the relationship between specimens of the same strain can be established, thus allowing us determine their evolution. In [11] a detection method that enforces carefully-chosen benign properties in trustworthy applications, but not in malware, is proposed. Another interesting approach is [40], where a scalable detection engine for inspection of APKs is designed using Multifeature Collaborative Decision Fusion (MCDF). In general terms, the static analysis has the advantages of simplicity at the data extraction stages and efficiency. However, it is a method liable to be deceived by obfuscation schemes. Furthermore, due to its lack of ability in defining the behavior of the applications at runtime, it often does not achieve high accuracy.

- *Dynamic analysis*. The analysis based on the study of dynamic features monitors the system behavior at execution, and gathers the activities of applications through various parameters. As referred in [16], the most common features on the dynamic analysis are the sequences of system calls involved in the execution of suspicious applications. A typical example of this is Crowdroid, an intrusion detection system proposed in [4] that considers their frequency of occurrence. In other approaches, such as [27] the Android malicious repackaged applications are identified by analyzing the thread-grained system call sequences. This group of proposal often carries out the extraction of the required data in isolated and secured environments, which are defined as sandboxes. In this way it is possible to observe the modus operandi of the attacks without compromising the protected system. In [38] a deep state of the art approach to sandboxing is presented. But the dynamic analysis also considers other features. For example, in [51] the behavior of the applications is modeled based on the requested permissions. In addition, the effectiveness of the implementation of metrics based on monitoring the power consumption of the devices is discussed in [20]. In [39] an intrusion detection system based on the recognition of anomalous behaviors on the activities of applications in networks is presented. In summary, the analysis of dynamic characteristics is very accurate. However it needs the use of a significant amount of computational resources, a situation that could be un-viable in some devices, hence requiring the availability of additional infrastructure.

- *Mixed analysis*. The more complex monitoring environments are usually more susceptible to become compromised. Because of this, it is common to combine both static and dynamic methods. This collaboration is referred as analysis based on mixed or hybrid data. A good example of this approach is in [48], where statistical analysis of the code and the *AndroidManifest.xml* is performed; in addition, different dynamic features are taken into account, such as registers, system calls or network traffic. Another interesting publication is [36], where the problem of the overload caused by dynamic analysis systems is reduced. With this purpose, a first study of the code of the analyzed applications is performed, looking for suspicious sentences. In [22] an intrusion detection system that adopts the criminal profiling methodology in the malware analysis is proposed. It is based on the analysis of heterogeneous data, among them opcodes, metadata in *AndroidManifest.xml*, certificates or the malicious behavior patterns of the application at runtime. Thereby, the dynamic analysis is simplified by focusing only on the activities arising from the labeled code, greatly reducing the amount of information to be taken into account. In general terms, hybrid proposals equilibrate benefits and drawbacks of the static and dynamic detectors, thus providing versatility in the adaptation to the various use cases.

- *Metadata analysis*. The last group of publications is based on metadata analysis. The metadata were defined in [16] as the information of the applications known prior to their download from the various stores. They provide different features, such as the requirements indicated by authors, opinions, comments, reputation, languages or their geographical distribution. A good example of the study of metadata is in [33], where information related with the software requirements displayed in the application stores is applied. For this purpose Natural Language Processing (NLP) techniques to identify sentences that explain the need for a given permission in an application description are implemented. In [45] a greater variety of information is taken into account. Depending on the platform, these include the description of the application, permissions, ratings, or information about the developers. The main advantage of the metadata analysis is that it is capable of recognize malware before it is downloaded. However, the success of these methods depends on easily manipulated information by the attackers. This could facilitate the evasion of the defensive schemes and lead to misclassifications.

## 3. Sequence Alignment

In nature, the crossover of individuals of the same species involve the appearance of different genetic alterations, such as variations on gens (substitutions), incorporation of new genes and chromosomes (insertions), or their omission (elimination). When these mutations pose advantages over other specimens on the same population, the individuals increase their survivability. This is the base of the natural selection. The comparison of the genetic similarities between the genotype of ancestors and offspring is studied in depth by bioinformatics. With this purpose, this area provides a great collection of strategies, which are able to preprocess, model and measure the various genetic features. The sequence alignment methods are part of this set. They originally determined the degree of similarity between strings of elements (usually DNA, RNA or proteins), which usually correspond to nucleotides or amino acids, and are represented by symbols from a finite alphabet $L$. A wide variety of sequence alignment algorithms has been proposed over the past years. By taking the number of sequences to be compared into account, they are divided into two methodologies: pairwise and multiple alignments. Pairwise sequence alignment methods are used to compare two query sequences. On the other hand, the multiple alignments extend the pairwise methods to analyze more than two sequences at a time, representing a significant increase in computational complexity. Note that these strategies also vary depending on the characteristics of the sample. Consequently, they require observing different aspects of the genotypes, such as their genes, structural features or their phylogenetic trees.

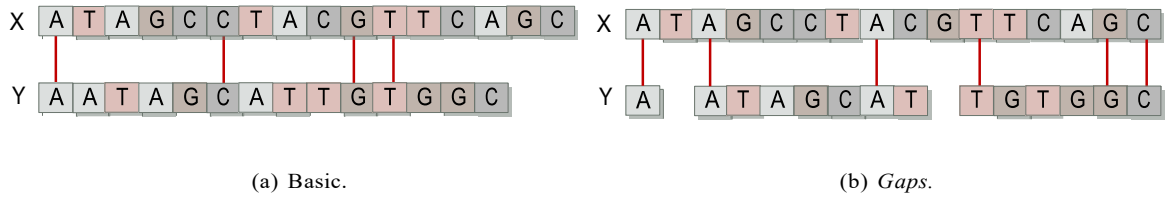

(a) Basic.  (b) *Gaps.*

Figure 1: Example of Sequence alignment.

Given that in our approach, only the pairwise alignment is applied, hereinafter only this method will be referred. In particular, the efforts carried out adopt the generalization to the problem of find the length of the Longest Common Subsequence (LCS) between two segments, according to the classical solution published in [47]. Therein different operations on the original sequences are performed, such as substitutions, insertions and elimination. They are achieved through the incorporation of new symbols into the chains, which are defined as *gaps*. The similarity of each pair of sequences in these configurations is measured by calculating the best distance between all their possible subsequences.

A practical example of this process is shown in Fig. 1a and Fig. 1b. In the first of them, two sequences are compared without alignment. Let the alphabet $L$ = {*A, C, T, G*} where each symbol represents an amino acid: Adenine (*A*), Cytosine (*C*), Thymine (*T*) and Guanine (*G*). The sequence *X* is {*ATAGCCTACGTTCAGC*} and the sequence *Y* is {*AATAGCATTGTGGC*}. There are four matches {*A*}, {*C*} and {*GT*}. In Fig. 1b the sequences are aligned by the insertion of a pair of *gaps*. This leads to recognizing a greater amount of matches, in particular six: {*A*}, {*A*}, {*A*}, {*T*} and {*GC*}. Assuming a simple heuristic that adds +1 to the score when a new match is found and not includes penalties, the score at the first example is 4 and the score after the alignment is 6.

The definition of appropriate scoring heuristics directly affects the results. At least they should consider matches, non-matches and *gaps*. In bioinformatics they are also built considering the class of amino acids involved in matches and non-matches, which often required specialized knowledge bases, as discussed in [25]. In addition, the location of *gaps* and the scoring heuristic usually depends directly on the alignment methods. These strategies have traditionally been based on the global, local or hybrid reallocation of their elements, as described below.

- *Global alignment*. This family of algorithms performs alignment considering the complete sequences to be compared. The most important method is well known as Needleman-Wunsch and was proposed in [31]. It usually yield better results when it is supposed that the sequences should be almost equal, as it provides an overview of their global features

- *Local alignment*. The algorithms based on local alignment aims on find the most similar regions within the sequences to be compared. They are commonly implemented as variations of the Smith-Waterman method proposed in [41]. It is based on combining all the partial global alignment scores of the various subsequences within the original data. Because of these characteristics, the local alignment provides high effectiveness when analyzing chains with different lengths.

- *Semi-global alignment*. The hybridization between global and local alignment often is referred as semi-global alignment. It is usually carried out by the implementation of slightly modifications on the Smith-Waterman method; highlight among them the no application of penalizations at the beginning and the end of sequences. They aim on compare the similarity between complete sequences and subsequences. Because of this, they are usually recommended for the comparison of chains with very different lengths [32].

The sequence alignment algorithms were original implemented by dynamic programming schemes. But manage large amounts of information, as is the case of the billions of nucleotides within sequences in nature, entailed their simplification by heuristic approaches, which reduce computational cost, but decrease accuracy. Most of these heuristic approaches are studied in [32].

Due that the analysis and comparison of sequences is a common problem in different fields of research, the sequence alignments has been used frequently in diverse use cases. For example, in [21] they are invoked to extract out the representative patterns which denote specific daily activities of a person from the training patterns. A very different use case is shown in [37], where sequence alignment is applied by web crawlers to detect and remove duplicate documents without fetching their contents. Despite their popularity, they are unusually considered in order to face the challenges on information security. In this case, they are focused on the recognition of actions performed by the users of the protected systems. A good example of their contribution is described in [8], where they are applied in the detection of masquerade attacks. This approach aligns sequences of system calls executed by the system users looking for impersonations. In addition, the effectiveness on the implementation of the different alignment algorithms is discussed.

## 4. Recognition of malware in Android

In this section the main features of the proposed detection system are described. Because of the many challenges inherent in the recognition of malware in mobile devices, prior to its development it is important to stress those aspects which have led to the definition of the design principles of the approach. In order to delimit the situations to be taken into account, the following enumerates the various assumptions which have been considered.

- Android is the most used operating system for mobile devices and 99% of the malware for smartphones is directed against this platform. The proposed intrusion detection system assumes that the detection of a larger portion of malware on this environment is possible. Thus, it is aimed to identifying and preventing the execution of malicious software on such devices.

- Given that an important part of Android malware is distributed from its application stores, this proposal assumes that it is possible their detection once downloaded, and before their installation on the victim systems.

- As it is mentioned in the related works, the execution of applications in an isolated and secure environment before they perform changes on the real system is possible. This proposal assumes that these methods are effective. In addition, it is based on them when extracting the information to be analyzed.

- In the bibliography, the accuracy of the dynamic analysis over other approaches is emphasized. As an important part of those publications, this proposal assumes that the dynamic analysis of the sequences of system calls invoked by the monitored applications is an effective measure for distinguish the malicious contents.

- Unlike many of the previous proposals, this approach assumes that in particular, by studying sequences of actions at the booting process of the applications, and hence taking into account their temporal relationship, it is possible to recognize malware. This assumption could be considered as the main null hypothesis of our research.

- This proposal also assumes that even when launched on the same device, it is difficult to find two equal booting action sequences of the same application. However, sequences from the same application present a significant resemblance, situation that allows distinguishing between legitimate applications, and modifications that contain infection vectors.

Bearing these in mind, the proposed malware detection system performs dynamic analysis of the behavior of the monitored applications, once downloaded from application stores. To this end, the system calls on their booting process are extracted and analyzed when executed in a secure and isolated sandbox. Thus the amount of information to be considered is reduced, since only the system calls related with their initialization are studied. Unlike similar approaches referred to in the bibliography, the temporal relationships of these actions are taken into account, and therefore the order they are initiated. In addition, by the implementation of a sequence alignment algorithm it is possible to identify the subsequences with greater similarity and determine their resemblance to the behavior of the legitimate applications. Upon completion of the analysis, only applications labeled as legitimate are allowed to make persistent changes on the mobile device.

The proper functioning of the proposal is defined by two main requirements. The first of them is to be able to accurately identify most of the analyzed threats. This also includes the issue of a low amount of errors when inspecting legitimate applications, thus reducing the penalization on the quality of experience of the user. On the other hand, the detection system must adapt to system characteristics. This involves operating with good performance, and low consumption of computational resources. Precisely because of the latter requirement, and taking into consideration the limitations on mobile devices, an important part of the analysis tasks are carried out by external infrastructure, as it is described in the following subsections. In general terms, these requirements meet an important part of the needs of the intrusion detection on smartphones. However there are other aspects that have been set aside, highlighting among them the fight against the various evasion strategies [29, 46], the interoperability with other security tools or adaptation to the various data protection policies. They are obviously important to grant its deployment on heterogeneous use cases. But they also add much more complexity to the problem to be solved, not being recommended for the better understanding of this first approach, but being a good aim for future work. The following describes the system architecture, monitoring, analysis and decision-making.

*4.1. Architecture*

The architecture of our proposal is distributed. This decision implies the addition of complexity to the design, but allows the distribution of the computational resources in different layers of data processing. Thus the tasks involving higher costs are performed on dedicated servers, while on mobile devices only behavioral patterns of applications downloaded by users are captured. Furthermore the protected devices must provide access to the network; a minor inconvenience considering the enhancement of their performance. As an alternative to the distributed scheme, the downloaded applications could include information related to the characteristics of their execution in metadata. These should vary according to the features of the device, such as the Android version or hardware. In this way the smartphone may complete the analysis by itself, assuming that it meets all the memory, battery and processing requirements. However the centralized approach has not been implemented in this work, being postponed to future research.

The implemented architecture is summarized in Fig. 2. Three processing layers are defined: monitoring, analysis and decision-making. The monitoring stage is the only one that takes place in the protected device.

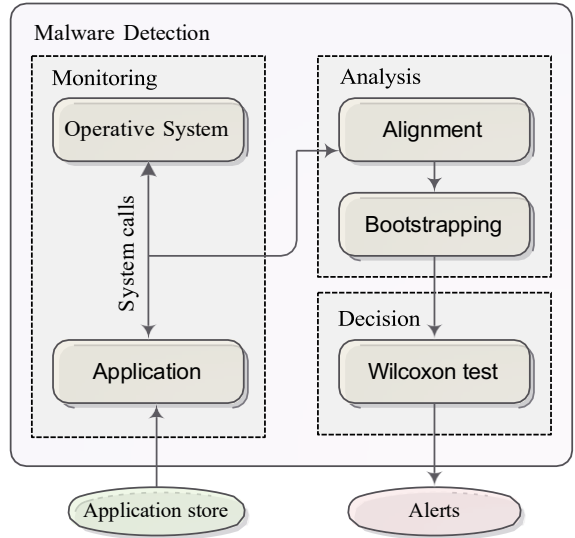

Figure 2: Architecture of the proposal.

Detection and decision-making occurs in a dedicated server. The behavior of the system is as follows: the suspicious software reach the protected device via application stores, then are installed and executed in an isolated and safe sandbox; therefore they are not able to make changes to the device, but their activity can be drawn. Under this environment, the sequences of system calls related with their execution are captured and transmitted to the analysis stage. This information is preprocessed and aligned with sequences of actions observed in their legitimate executions, shared by trusted users in similar devices. The resulting scores allow decision-making level determine their nature. Thus, if it is labeled as a potential threat, changes in the protected system are not made.

*4.2. Monitoring*

At the monitoring stage, the device captures in a sandbox the system calls launched by the suspicious applications at their boot processes. As in the previous works, the monitoring of the executed applications is performed by the diagnostic tool *strace* [4], which is present in most of the GNU/Linux systems, such as Android. This utility facilitates registering the system calls carried out by a program or process by audition of the system call interface, which allows communication between the kernel and the upper layers of the operating system. In order to study all the activities of a particular program from the time of its initialization, including those processes derived from it, the parent process (a.k.a Zygote) must be monitored. Zygote is a daemon triggered by the Android *init* process, responsible for generating all the application processes. In Fig. 3 the Android boot sequence and the role of Zygote are summarized. When an application launches Zygote, it creates the first Dalvik virtual machine and calls its main method, which preloads all necessary Java classes and resources, starts System Server and opens a socket to listen for requests of the starting applications. The monitoring stage implements *strace* in order to capture the sequences of system cells derived from zygote, after the suspicious application is launched. Note that aiming on reducing noise, the sandbox only executes the application to be analyzed.

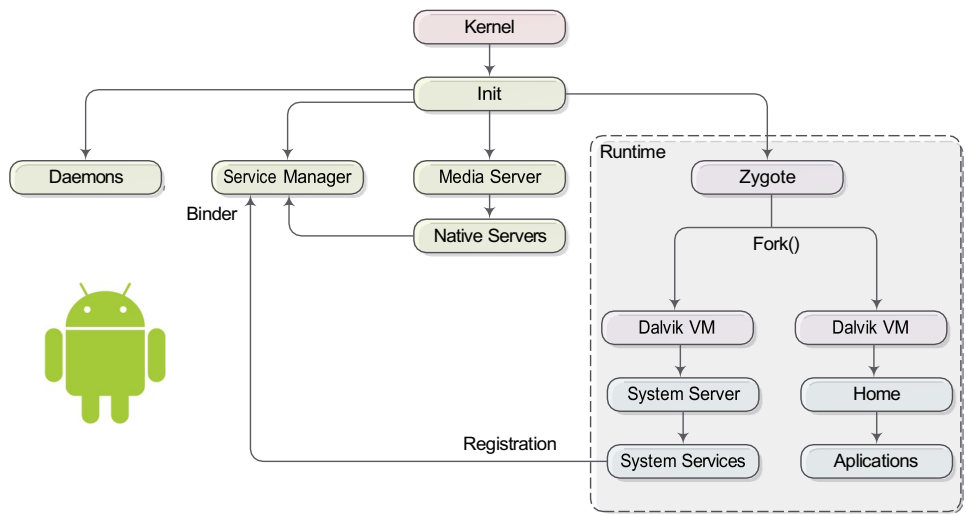

Figure 3: Android boot sequence.

On the other hand, and with the purpose of bypass the inherent characteristics of the device, the consecutive repetitions of the same actions are simplified in one of them. This further reduces the problems related with errors in the capture tasks, which are usually generated by the redundant data added by the monitoring software. It should also be taken into account that the length of the boot sequence may vary, often having around 2,000 system calls in small applications, and more than 50,000 in the most complex. To simplify treatment, they are preprocessed, so every type of action is associated with a symbol. Hence, if the operating system offers a repertoire of $l$ system calls, the alphabet $L$ that identifies every possible action has length $l$. Ideally, the complete sequences are studied. However, it is possible that because of computing limitations, only the first boot steps are aligned.

All captured sequences are transferred to a dedicated server, where they are compared with the legitimate executions shared by other users. If when analyzing, the application is labeled as legitimate, the new sequences are appended to the database. But only if it is verified that the applications are really clean, their sequences may be used as training samples to evaluate future executions. This prevents the attacker compromising the detection system by poisoning the set of reference samples.

*4.3. Analysis*

The main purpose of the analysis module is generating the metrics required by the system that allows a decision to be made regarding the nature of the suspicious software. The basic components of these metrics are the scores obtained by aligning the sequence of system calls related with the boot process of the suspicious application, with sequences corresponding with the launch of the legitimate application on similar mobile devices. The implemented sequence alignment algorithm is an adaptation of the global alignment method proposed by Needleman-Wunsch [31]. It is a dynamic programming scheme which correlate both sequences from a matrix $F$. The decision of implementing Needleman-Wunsch has been addressed keeping in mind that because of computational issues, the maximum length of the monitored sequences is defined. The captured activities often reach such amount, so the alignment algorithm usually operates at the case where both sequences exhibit the same dimension, situation that encourages the use of the global alignment paradigm. This choice is also supported by the fact that the sequences to be compared contains patterns of a similar priori.

Let the pair of sequences to be aligned $A$ and $B$, such as $|A| = m$, $|B| = n$, the matrix of their similarity $F$ has dimension $m \times n$. Note that given that the maximum length of the boot sequences is $l$, $F$ has dimension $l \times l$ in the worst case. Hence if $l$ is too large, lot of memory for storage is required. This potentially may become a problem if the proposal is not implemented in a distributed architecture. Even in this case, it could be an issue, because it needs the transmission of a large amount of information to the server-side, in this way consuming more of the network bandwidth contracted by the user. But obviously the greater the length of the sequences, the more information provide, allowing make a more accurate analysis. Consequently, it is recommended to select lengths as large as possible, according to the computational limitations of the deployment environment.

Within $F$ matrix, the value of each cell $F(i, j)$, $0 < i \leq m$ and $0 < j \leq n$, is the score of the best alignment between sub-segments of the first $i$ elements of $A$, and the first $j$ elements of $B$. Therefore $F$ stores the best alignment between the two chains, and $F(m, n)$ contains the optimal alignment. The matrix is recursively constructed from the following base cases:

$$F(0, j) = d \times j; F(i, 0) = d \times j \tag{1}$$

where $d$ is the *gap* penalty. From these data it is possible to begin filling the rest of the matrix. The value of each cell $F(i, j)$ is determined by its row, column, or diagonal, defined as the following expressions:

$$F(i, j) = max\{F(i-1, j) + d, F(i, j-1) + d, F(i-1, j-1) + S(A_i, B_j)\} \tag{2}$$

where $S(A_i, B_j)$ indicates the similarity between the elements $A_i$ and $B_i$. This value is often defined as a scoring matrix, which is the heuristic implemented by the algorithm. The computational cost of this process is $\Theta(n^2)$. This is one of the main reasons that led to use heuristic approaches when analyzing large sequences.

The alignment aims to match representative sets of system calls in a tested block with similar groups in the sequence derived from the suspicious application. It is a very similar situation to that discussed in [7], so an important part of the guidelines that have led to their resolution are taken into account. For example, *gaps* inserted in the sequence to be analyzed are differently penalized than those on the reference executions. In particular, *gaps* in the latter are slightly more penalized, as it is considered less desirable altering the order of execution of system calls in the reference samples (whose legitimacy is known a priori) than the order of the sequence to be tested. On the other hands, matches increase the score corresponding to the similarity between both sequences. As is usual in the bibliography, in the proposal mismatches do not influence the score. In related works this only occurs on certain contexts, and assuming particular knowledge bases [25], which is out of scope. With this in mind, in the analysis stage the following scoring function is considered:

$$S(A_i, B_j) = \begin{cases} 1 & \text{if } xA_i = B_j \\ 0 & \text{if } xA_i \neq B_j \end{cases} \quad (3)$$

where every single match scores +1, mismatches do not penalize, and in addition to $S(A_i, B_j)$, *gaps* are penalized non uniformly by assigning the value $d = -2$ if they are in the sequence to be evaluated, and $d = -3$ in the case of taking part of the reference samples.

Once $F$ is completely filled, the ordering of the sequences and insertion of gaps starts in the position $F(m, n)$. The matrix is traversed selecting as next position the best value of $F(i-1, j)$, $F(i, j-1)$ and $F(i-1, j-1)$. Thus the optimal solution at each position $(i, j)$ is considered. If the chosen value is $F(i-1, j)$ or $(i, j-1)$, then $A_i$ is aligned with a *gap*. But in the case of $F(i-1, j-1)$, it is aligned with $B_i$.

In Fig. 4 an example of $F$ matrix built with the Needleman-Wunsch method is shown. There two sequences are aligned: *{ABCDEBE}* and *{DEBFBCFDEE}*. The alphabet is defined as $\Sigma = \{A, B, C, D, E, F\}$ where each symbol represents a different system call: *create module* (A), *close* (B), *writev* (C), *waitid* (D), *execv* (E) and *fallocate* (G). At the end of the algorithm, the final score is 1, as is indicated in $F(m, n)$. The best alignment is calculated traversing the matrix following the best score of $F(i-1, j)$, $F(i, j-1)$ and $F(i-1, j-1)$, assuming that the current position is $F(i, j)$. By this way the sequence *{--A-BC-DEBE}* is the most similar to *{DEBFBCFDEE}*. From this figure, it can also be clearly seen how the matrix was built. For example, corresponding with the recursive expressions of the method, the position $F(1, 2)$ contains the best score of $F(0, 2) + d$, $F(1, 1) + d$ and $F(0, 1) + S(B_i, D_j)$. Given that the *gaps* are penalized ($d = -8$), the mismatch between B and D is also punctuated ($S(B_i, D_j) = -2$) and the related values on the matrix are $F(0, 1) = -8$, $F(1, 1) = -2$ and $F(0, 2) = -16$, then $max\{-10, -10, -24\} = -10$, concluding with $F(1, 2) = -10$, as is indicated in the appropriate cell.

Finally, it is important to characterize the set of reference sequences. The sequence to be tested gets a score from its alignment with each of the sequences within this collection, and at the end of the analysis stage, a vector $s$ of length n that summarizes all the calculated scores is built. Therefore the value of $n$ depends on the size of the set of legitimate executions considered as reference. The configuration of this set is relevant, and corresponds to the problem of validating models common in the various machine learning algorithms. The set can be seen as the model of legitimate execution of the application. Because of this it is easy to realize that the more representative is the dataset, the more accurate analysis. Furthermore, the smaller is the set, the faster the algorithm.

|   | D | E | B | F | B | C | F | C | E | E |
|---|---|---|---|---|---|---|---|---|---|---|
|   | 0 | -8 | -16 | -24 | -32 | -40 | -48 | -56 | -64 | -72 | -80 |
| A | -8 | -2 | -9 | -17 | -25 | -33 | -41 | -49 | -57 | -65 | -73 |
| B | -16 | -10 | -3 | -4 | -12 | -20 | -28 | -36 | -44 | -52 | -60 |
| C | -24 | -18 | -11 | -6 | -7 | -15 | -5 | -13 | -21 | -29 | -37 |
| D | -32 | -14 | -18 | -13 | -8 | -9 | -13 | -7 | -3 | -11 | -19 |
| E | -40 | -22 | -8 | -16 | -16 | -9 | -12 | -15 | -7 | 3 | -5 |
| B | -48 | -30 | -16 | -3 | -11 | -11 | -12 | -12 | -15 | -5 | 2 |
| E | -56 | -38 | -24 | -11 | -6 | -12 | -14 | -15 | -12 | -9 | 1 |

Figure 4: Example of alignment with Needleman-Wunsch.

In the implementation, the amount of executions taken as reference is predefined. It is assumed that the number of available legitimate sequences is greater or equal to the set applied in the analysis. Because of the construction of different reference sets, the classifications could vary, situation that derives on irregular behaviors of the system. Fortunately, this is a problem widely discussed by the machine learning community [24]. Among the various existing solutions, Bootstrapping aggregating was implemented [3]. The main goal of this method is building $m$ training datasets $C_i$ bagging from a larger training set $C$, uniformly and with replacement. In this way the system generates different datasets from the original group of sequences of legitimate executions, and then uses each of them on a different instance of the classifier. In particular, all the possible combinations of n legitimate executions are considered. From these analysis, $m$ scoring vectors $S_i$, $0 < i \leq m$ are obtained. Then the results are pooled, and from them the final classification provided by the detector is decided. The aggregation of scores is performed by arranging the elements of each $S_i$ from lowest to highest. The value of every position $j$ on the final vector is the average of the values in $j$ of all the $S_i$. Because of this, at the end each element on the final scoring vector contains the average score of the partial results in the same position.

In Fig. 5 the management of the dataset of initializations sequences when analyzing suspicious applications is summarized. When a new request of analysis reaches the intrusion detection system at server-side, the subsets considered by the instances of the detector are bagged. Consequently, the different analyses are carried out in concurrency. Each of them provides a particular scoring vector, thus representing the similarity of the tested sequence with all the samples within the reference sets. Finally, the elements on the scoring vectors are sorted. This information is pooled in a final vector by calculating the average score of each position. In this way the metric for the decision-making stage is generated.

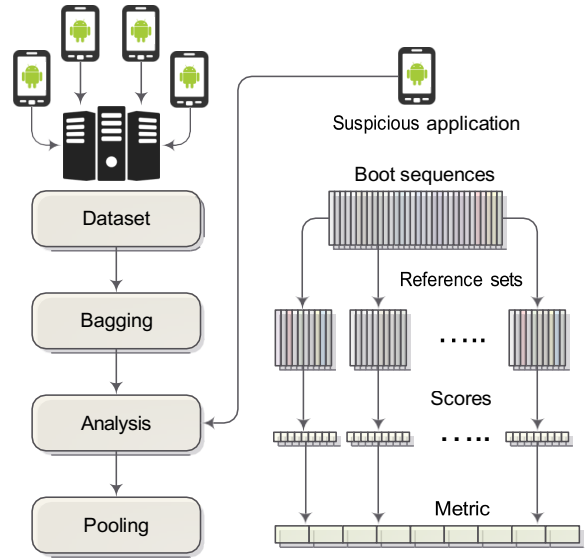

Figure 5: Bootstrap aggregating when analyzing applications.

The main drawback of bagging reference sets is that computationally, involves a very expensive strategy, so it is not suited for centralized architectures. Moreover, its scalability is limited, so if the database grows too much, the method would lack efficiency. A common way to reduce this problem is by randomly selecting the group of samples to be divided into the training sets. But in order to avoid randomness, the number of samples for each application in the database is fixed. This could affect the process of upgrading $C$ by adding the latest sequences tagged as legitimate and verified. To reduce this problem, the database is updated each time a new sequence is added, but replacing some of the older entries. In this way it also preserves consistency.

*4.4. Decision-Making*

At the decision-making stage, the suspicious applications are labeled as legitimate or malicious considering the information provided by the analysis module. In the first case, they are allowed to act on the protected environment, and their boot sequences may be added to the dataset. But if they are malware, the intrusion detection system reports the incidences, and the potential threats are blocked.

In particular, the labeling step has as input, the vector of scores obtained after aligning the sequence of system calls gathered by monitoring the Zygote process at the boot of the suspicious applications, with the bootstrapped sets of references. Apart from these data, it also considers the vector of scores obtained after aligning all the legitimate sequences with each of them. With all this information the Wilcoxon signed-rank test is performed [49].

Wilcoxon signed-rank test is a non-parametric test that operates on vectors with paired elements, comparing those with the same index. To this end it is assumed that the vectors contain $n$ pairs of observations, each of them referred as $(x_i, y_i)$, $0 < i, \leq n$. Other assumptions are that the data come from the same population, each pair is chosen randomly and independently, and data are measured at least on an ordinal scale (cannot be nominal). The choice of this method is justified by two premises. First, it is not possible to assume the distribution of the scoring vectors, which leads to apply a non-parametric test. This reduces the search space, but there are still many evaluation schemes framed within this family. Nevertheless, and unlike the other non-parametric tests (such as the Mann-Whitney U-test), the samples are not independent, leading us to treat data pairs (analogue of dependent t-test for paired samples), where the Wilcoxon signed-rank test is the best solution.

The objective of the test is to determine that the values $x_i$ and $y_i$ are equivalent. In the proposed system, these coincide with the elements $j,i$ of the scoring vectors to be compared. The null hypothesis is that the median difference between pairs of scores is zero. The test statistic is $W$, defined as the smaller of $W_+$ (sum of the positive ranks) and $W_-$ (sum of the negative ranks). Bearing this in mind, when the null hypothesis is satisfied, it is expected to find similar numbers of lower and higher ranks that are both positive and negative, so $W_+$ is close to $W_+$.

In order to verify this assumption, each paired difference, $d_i = x_i - y_i$ is calculated. Every $d_i$ is ranked ignoring the sign (i.e. assign rank 1 to the smallest $|d_i|$, rank 2 to the next, etc.). Then they are tagged according to their sign. The sum of the ranks with positive differences $W_+$ and negative differences $W_-$ are calculated. With this, the statistic $W$ is calculated according to the following expression:

$$W = min\{W_+, W_-\} \quad (4)$$

From the statistical $W$ it is possible to calculate a p-value. In case of non-representative $n$ (usually $n < 20$), this implies take into account the table of critical values for the Wilcoxon Signed Ranks test described in [49]. If $n$ is large enough, it is calculated based on the normal distribution, as the following expression summarizes:

$$Z = \frac{\frac{W - n(n+1)}{4}}{\frac{n(n+1)(2n+1)}{2}} \quad (5)$$

The test is passed if $p < I$, where $I$ is the confidence interval defined for the evaluation. This is interpreted as the difference between the populations is significant, and therefore it is not due to randomness. Consequently, it is considered that the sequence of system calls to be analyzed does not resemble the executions of legitimate applications within the reference sets, hence being tagged as anomalous. In this case a potential malicious application is detected.

## 5. Experimentation

In the experimentation, the proposed system has been deployed on different mobile devices with various versions of the Android operating system. As no significant differences have been found in effectiveness, the device setting is not considered in the group of sensitivity parameters of the test. The analyzed collection of applications combines legitimate and malicious specimens of the public datasets Genome [52] and Drebin [2]. The following 19 legitimate applications were applied: *Diner Dash 2*, *Jaro*, *Mash*, *Plumber*, *Fruits Matching*, *Scrambled Net*, *Solitaire*, *Tap and Furious*, *Robotic Space Rock*, *Basketball shot*, *Monkey Jump 2*, *Whites out*, *Super touch down*, *Tilt Mazes*, *Helix*, *DailyMoney*, *Sanity*, *Best Voice Changer* and *Z-test*. Their malicious versions contain infection vectors of the following 9 specimens: *DroidKungFu*, *Plankton*, *Geinimi*, *GinMaster*, *Cogos*, *jSMSHider*, *VdLoader*, *Gapev* and *Gamex*.

The monitoring was distributed at different intervals throughout a time period of 2 months. A total of 300 different boot sequences per application were considered (150 legitimate and 150 malicious, 3 per device) assuming variations in mobile devices and operative system. So the complete dataset contains 2850 executions of the legitimate applications and 2850 sequences of attacks. It is worth mentioning that all the data gathering tasks were performed in real mobile devices owned by different users. Because of the obvious time consumption of them, it was not possible to consider a much larger selection of applications.

The evaluation methodology involves the study of the behavior of the proposed strategy when analyzing the previous malicious and legitimate applications. In the first case, the intrusion detection system adopts the legitimate collection of samples as reference set, and the malicious collection as test set. Then the malware is dissected one by one, and the results are observed. On the other hand, in order to measure the sensitivity of the approach to report false positives, the legitimate samples are needed in both, reference and test sets. With the purpose of improving the use of the dataset and avoid collisions (i.e. occurrences of an identical sample within the reference and test sets, on a particular configuration), the experimentation with only legitimate samples implemented a cross-validation scheme. The different boot sequences related with the execution of each application were divided into three disjointed groups: *A*, *B*, and *C*. From them three different scoring vectors were generated ($V_1$, $V_2$, $V_3$):

- $V_1$ contains the scores obtained by aligning all the samples within *A* and *B*.

- $V_2$ contains the scores obtained by aligning all the samples within *A* and *C*.

- $V_3$ contains the scores obtained by aligning all the samples within *B* and *C*.

Summarizing, each vector contains the scores required when analyzing the remaining group. For example, to calculate the false positive rate when studying boot sequences within group $C$, the scoring vector $V_1$ is taken as reference. This is because it is built from legitimate samples within $A$ and $B$, thus taking advantage of the complete dataset. Notice that for example, in this case it would not be honest to analyze sequences within $A$, which already were considered in the training set. The evaluation of the proposal is divided into three stages:

1. *In-depth study of a particular use case.* The first test is an example that aims on facilitating the comprehension of the proposed method, and the importance of the sequence preprocessing tasks. During its progress, the application *DailyMoney* is analyzed, in both legitimate and malicious executions. There, the results obtained by aligning the various boot sequences are studied, and the relationships between the score vectors are discussed. Although the study focuses on a single application, the results are very similar to those of the others, plainly illustrating the problem surrounding the data gathering process. In addition, the advantages on preprocessing information are shown.

2. *Variations on the confidence interval.* In this experiment the accuracy of the proposal in function of the rigor with which decisions are made, is studied. All the samples are analyzed, and the TPR and FPR were calculated for each adjustment value. It is expected that the higher is the confidence interval, the lower the True Positive Rate (TPR) and False Positive Rate (FPR). The confidence interval that minimizes precision errors is applied for the remainder of the experimentation.

3. *Variations on the length of the boot sequences.* In this test the behavior of the approach when modifying the length of the boot sequences is measured. Its interest lies in the possibility that when dealing with short sequences, the boot activities of the applications are not completely represented in the captures. This situation makes the detection tasks difficult, because the payload of the threat could not be represented. In the opposite scenario, if the sequences are too long they could include redundant information, which also worsens the quality of the decisions. At this experiment all the samples within the dataset were analyzed, and the TPR and FPR were calculated and discussed for each length.

The proposed detection system behaves satisfactorily if after the test met the requirements extensively described in section IV: accuracy when identifying threats, report low amount of detection errors when inspecting legitimate applications and adaptation to the protected environment characteristics.

## 6. Results

*6.1. In-depth study of a particular use case*

For this experiment, a vector of reference scores has been built considering 30 samples of executions of the application *DailyMoney*. The length of their boot sequences is 2,500 system calls, and the confidence interval of the Wilcoxon signed-rank test is $I = 0.001$. In a first step of evaluation, the preprocessing of data previously described in section IV (i.e. reducing the consecutive repetitions of system calls into one action) was not implemented. The various alignments performed for the construction of the vector of reference scores are shown in Fig. 6a. There, rows and columns indicate the legitimate boot sequences of the reference sample set, so each cell contains the score on each partial alignment. For ease their understanding, the scores are displayed according to the color chart which is shown on the right illustration. On the other hand, in Fig. 6b a similar structure is shown, but this time the rows that indicate the boot sequences of the same application include harmful payload. To the naked eye it is possible to realize that the difference in scores is fairly representative: the average score in the first structure approaches 1248.9 while in the second is 1033.4. With this scheme it was possible to successfully detect 100% of the inspected malicious samples. However, the accuracy obtained when analyzing legitimate applications was not as satisfactory; the FPR is close to 32%, thus hindering their deployment in real environments.

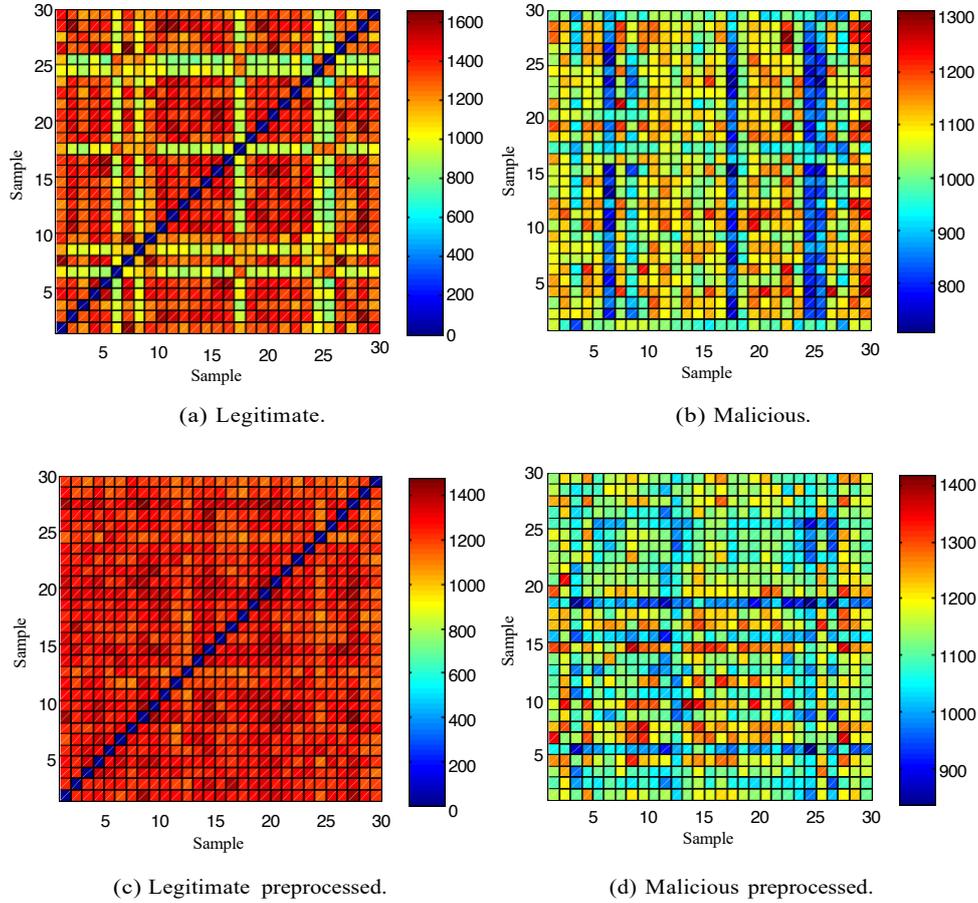

Figure 6: Alignment scores in *DailyMoney*.

Although the example is shown for a specific application, the false positive problem has persisted through-out the experimentation. It only has been reduced by preprocessing. In this way a large part of the observed noise is mitigated during the capture tasks. The method led to discover important differences between sequences from different devices, but easily offset. To illustrate this, in Fig. 6c the scores of the alignment of legitimate sequences (analogous to Fig. 6a) are shown, and in Fig. 6d the malicious executions (similar to Fig. 6b) are displayed, all of them after applying preprocessing. At first glance, it can be observed as in the legitimate matrix of the latter case, the darker colors dominate (and hence the higher scores). In general terms, the scores are higher and with less variation, reaching an average score of 1431.6. In addition, for the harmful boot sequences the results are also lower, averaging 829.6 In this case the TPR approaches 98% and FPR is close to 2%. Therefore their difference of means is greater, facilitating the decision-making orchestrated by the Wilcoxon test.

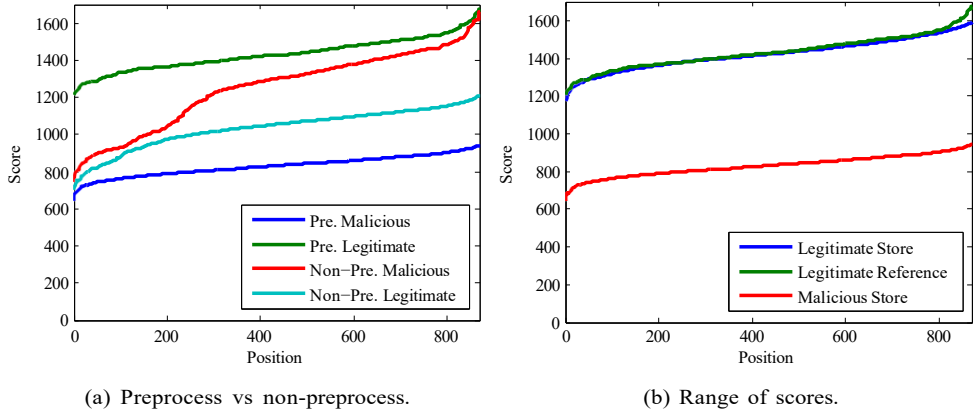

(a) Preprocess vs non-preprocess.

(b) Range of scores.

Figure 7: Score vectors in *DailyMoney*.

The increases of the paired scores after ordering their mean values are shown in Fig. 7a. As can be seen, after preprocessing the score vectors pose a more homogeneous distance, which less frequently may lead to labeling mistakes. But without preprocessing, a significant convergence region appeared. Its location has changed along the experimentation depending on the application.

Finally, in Fig. 7b different vectors of average and sorted scores, related to the analysis of *DailyMoney* are shown: the vector of reference scores built during training (*Legitimate Reference*), the vector of mean scores obtained by analyzing instances of the legitimate application once download from the Application Store (*Legitimate Store*) and the vector of average scores obtained when analyzing compromised instances (*Malicious Store*). There the similarity between the first two and their significant distance with the third vector is obvious.

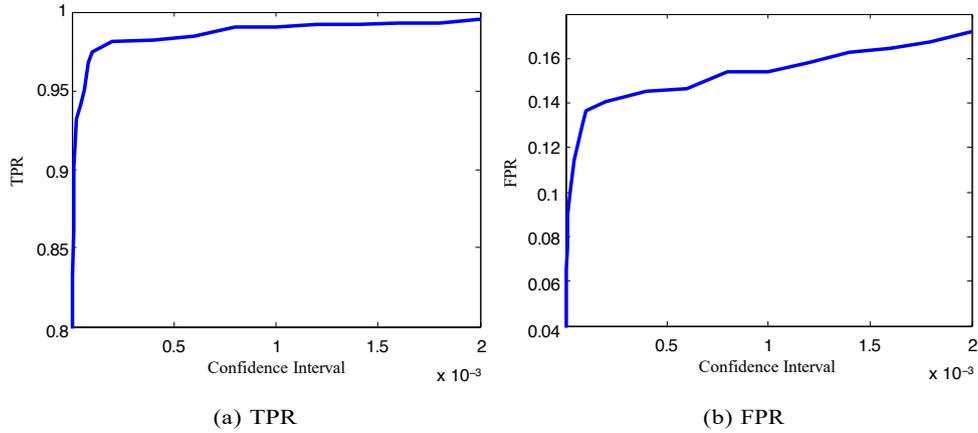

(a) TPR  (b) FPR

Figure 8: Average variations on confidence interval

## 6.2. Variations on the confidence interval

At this test, the system accuracy has been measured by analyzing its reactions to variations on the confidence interval in range 0.002-0.0000004 and assuming a fixed sequence length of 1,000 system calls. In Fig. 8a the resulting average hit rates are shown. When the confidence interval has lowest error tolerance (i.e. close to cero), the TPR value is nethermost. In particular, at the displayed range the worst case implies approximate TPR of 80%. On the other hand, the higher is the error tolerance, the greater precision. The system behaves with better accuracy when TPR is near 99.6%.

On the other hand, in Fig. 8b the relationship between false positives and the value of the confidence interval is shown. The lower is the error tolerance, the better the result. So when the interval confidence is close to zero, the FPR approximate 4%. Analogously, the higher is the error tolerance; the worst is the system behavior. The worse observed TPR is close to 17.19%. At this test, the optimal balance between TPR and FPR is obtained when the confidence interval approaches 0.03, where the TPR is 95.01% and the FPR is 10.03%. This demonstrates that besides the confidence interval, other features should be taken into account in order to reduce the false positive rate.

## 6.3. Variations on the length of the boot sequences

With the aim of determine the impact of the sequence length in system accuracy, the range of 50-2500 system calls per boot was analyzed. In Fig. 9a the dependence of the obtained precision to this parameter is described by the following three curves: the best hit rate per length (*Max TPR*), the average accuracy (*Average TPR*) and the worst cases (*Min TPR*). There, it can be seen that the detection system requires monitoring a minimum number of actions in order to work properly. In particular, in *Max TPR* this occurs from 500 system calls, and the hit rate approaches 100%. In the case of *Average TPR* it happens after 750 actions, with 95.37% accuracy. Finally, in *Min TPR* stability is observed from 1400 system calls, with precision of 59.33%. Despite the fact that *Min TPR* presented discouraging results, these occur very infrequently. Indeed, the average difference between *Max TPR* and *Average TPR* is less representative (approximately 0.14427) than that observed between *Average TPR* and *Min TPR* (about 0.35785). In general terms, the trend observed throughout the experiment has much to do with the sequence alignment algorithm and the distribution of the malicious contents on the boot processes of the applications. The payload of the attack is not released until the beginning of the basic functionality of the application, situation that affects the required amount of actions to be monitored in order to success. Only then will the alignment algorithm be capable of identifying regions of divergence between sequences, thus allowing the presence of anomalies to be reported.

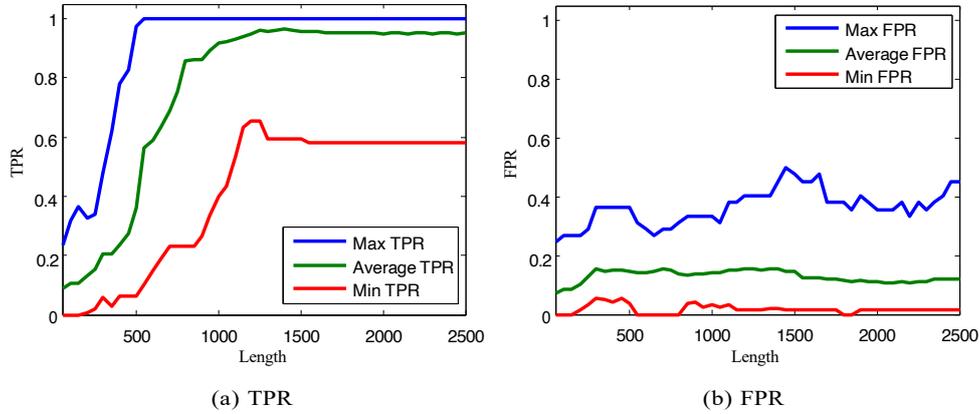

(a) TPR        (b) FPR

Figure 9: Average variations on length

Similarly, in Fig. 9b the relationship between the false positive rate and the length of sequences is shown. Unlike the previous experiment, from the outset regularity has been observed (except for *Max FPR*). *Min FPR* is stabilized before analyzing the first 500 system calls, with a false positive rate of 1.7%. *Average FPR* is also stabilized before the first 500 activities; however its false positive rate is close to 11.7%. Finally, *Max FPR* displays less regularity. Its values vary continuously in an interval ranging from 24.44% to 47.61%. As in the previous test, *Average FPR* has greater similarity to the optimal values, but in this case they are described by *Min FPR*; their average difference is 0.1105. However its mean difference with *Max FPR* is considerably higher, approaching 0.2356. The early saturation of these curves is mainly due to the lack of divergence regions within the monitored sequences, contrary to what occurs when they include malicious content. Therefore, differences between sequences are global and constant throughout the boot process. They are mainly due to the noise caused by the context in which the applications are launched on the sandbox.

In conclusion, both confidence interval and sequence length have a direct impact on the system sensitivity. Because of this, they must adjust to each device at every use case. Table 1 summarizes the final results for each of the analyzed applications, once stabilized the detection capability of the sensor and with sequences of length 2000. The average hit rate when studying applications with malicious content is 98.61%, and the average false positive rate is 6.88%.

| Application | TPR | FPR |
|---|---|---|
| Diner Dash 2 | 1 | 0.066 |
| Jaro | 1 | 0.24444 |
| Mash | 1 | 0.08696 |
| Plumber | 0.99333 | 0.01887 |
| Fruits Maching | 0.85333 | 0.02609 |
| Scrambled Net | 1 | 0.10638 |
| Solitaire | 1 | 0.0375 |
| Tap and Furious | 1 | 0.08333 |
| Robotic Space Rock | 1 | 0.10714 |
| Basketball Shot | 1 | 0.01739 |
| Monkey Jump 2 | 1 | 0.09565 |
| Whites Out | 1 | 0.16667 |
| Super Touch Down | 1 | 0.13954 |
| Tilt Mazes | 1 | 0 |
| Helix | 1 | 0.01770 |
| DailyMoney | 1 | 0.02222 |
| Sanity | 1 | 0.02321 |
| Best Voice Changer | 0.89 | 0 |
| Z-Test | 1 | 0.05000 |

Table 1: TPR and FPR per application with length 2000

*6.4. Discussion and comparison*

Keeping the results obtained in mind, the following items are highlighted:

1. If the system is properly configured, the results in terms of accuracy demonstrate high precision when dealing with malicious applications (TPR=0.9861), and the false positive rate is 6.88%.

2. When preprocessing the score vectors, they pose a more homogeneous distance, which less frequently may lead to labeling mistakes.

3. The lower is the error tolerance of the approach (i.e. lower confidence interval), the better the result in terms of identification of malicious samples. But this penalize the false positive rate.

4. The length of the sequences should be large enough to capture the entire startup activities performed by the suspicious applications. If it must be fixed because of computational limitations, the larger sequences, the greater true positive rate. The false positive rate is independent of the length, and it has much to do with the noise generated during the extraction of the sequences.

These statements demonstrate that by studying sequences of actions at the booting process of the applications, and hence taking into account their temporal relationship, it is possible to recognize malware. The proposed pattern recognition method has proven effective in this task (1). But it has demonstrated sensitivity to the adjustment parameters (3) and how the data is collected for analysis (2)(4). On the other hand, the comparison of the observed accuracy with the effectiveness of the most precise publications of the bibliography, states that our capability of recognition of malware is similar. This is a very important point considering that our proposal requires much less information, and that its search space is reduced to only the startup actions of the applications. However, the main disadvantage on the results is a trend to issue a higher rate of false positives that the proposals that perform the analysis of the complete actions [42, 15], were the result ranges on 1-3%. It is noteworthy to mention that our method does not compete directly with these publications. This paper introduces a novel first line of defense based on analyzing only boot sequences, which pose a different monitoring scenario, with a priori less likely to recognize malware. Therefore it is perfectly able to complement other general purpose schemes.

From an empirical point of view, our experimentation allow one to be optimistic about the effectiveness of the approach. On theoretical grounds, other interesting details should be taken into account. For instance, the strategy developed inherits the characteristics of the sequence alignment algorithms. Because of this it is capable of identify high-level patterns within the alignment elements and the sheer number of parameters that can be changed to suit different types of data [8]. This is a relatively unexplored field to address malware recognition on mobile devices, which fits very well with the characteristics of the data that is usually studied. It also should be borne in mind that in particular, this paper takes advantage of the global alignment methods, which usually yield better results when it is supposed that the sequences should be almost equal, thus providing an overview of the startup features. In return, the proposal is penalized in terms of resource consumption: the Needleman-Wunsch is implemented by dynamic programming, and hence the memory of the system must store large data arrays. The problem of limited resources is present throughout the bibliography, and in our approach is addressed in the same way as in most of the related works: by performing tasks involving higher costs on dedicated servers [4, 44]. This implies that the protected devices must provide access to the network; a minor inconvenience considering that the programs to be analyzed are downloaded from application stores, and that the extracted information is preprocessed and aligned with sequences of actions steadily updated by other trusted users, both of them involving connectivity.

## 7. Conclusions

A novel intrusion detection system for recognition of malware on mobile devices, specialized in the Android platform, has been proposed. Its main goal is to prevent the installation of the malicious software on the victim system, once downloaded from the application stores. To do this, the boot sequences of system calls performed by the suspicious applications are studied in deep. It involves an elaborate treatment process that includes the capture of system call sequences in an isolated and secure environment, their analysis by sequence alignment methods, and the decision of their nature by statistical hypothesis testing.

The experimentation performed three different tests, and the analyzed samples of both, legitimate and malicious applications, were provided by the datasets Genome and Drebin. The results obtained demonstrates that studying boot sequences is a lightweight and efficient complement to the conventional dynamic analysis schemes, posing an unused first line of defense.

However, the proposal also showed sensitivity to changes in their adjustment parameters (i.e. sequence length and confidence interval), situation that could derive in future modifications to the original detection strategy, such as variations on the sequence alignment algorithm, scoring system, or the decision-making method. In addition, along with the article, different alternatives for future research have been proposed, including, straightening against adversarial attacks, the implementation of a centralized version of the approach, or the discussion on the information about the boot of applications is shared, their reputation system or how to deal with the recent privacy issues.

## Acknowledgements

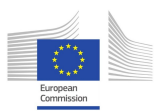 This work was funded by the European Commission Horizon 2020 Programme under Grant Agreement number H2020-FCT-2015/700326-RAMSES (Internet Forensic Platform for Tracking the Money Flow of Financially-Motivated Malware).